\documentclass[12pt,preprint]{aastex}
\usepackage{enumerate}

\newcommand{\Mearth}{$M_\oplus$}
\newcommand{\Msun}{$M_\odot$}

\newcommand{\AU}{{\sc au}}

\newcommand{\Corot}{\emph{CoRoT}}
\newcommand{\Kepler}{\emph{Kepler}}

\slugcomment{}

\shorttitle{Ices in Low-Mass Extrasolar Planets}

\shortauthors{Marb\oe uf et al.}

\begin{document}

\title{Composition of Ices in Low-Mass Extrasolar Planets}

\author{
U.~Marboeuf\altaffilmark{1,7},
O.~Mousis\altaffilmark{1,7},
D.~Ehrenreich\altaffilmark{2},
Y.~Alibert\altaffilmark{1,3},
A.~Cassan\altaffilmark{4,7},
V.~Wakelam\altaffilmark{5},
\& J.-P.~Beaulieu\altaffilmark{6,7}
}

\altaffiltext{1}{Institut UTINAM, CNRS (UMR 6213), Observatoire de
Besan\c{c}on, France; ulysse.marboeuf@obs-besancon.fr, olivier.mousis@obs-besancon.fr, yann.alibert@obs-besancon.fr}

\altaffiltext{2}{Laboratoire d'Astrophysique de l'Observatoire de Grenoble, Universit\'e Joseph Fourier, 
CNRS (UMR 5571), France; david.ehrenreich@obs.ujf-grenoble.fr}

\altaffiltext{3}{Physikalisches Institut, University of Bern, Switzerland}

\altaffiltext{4}{Astronomisches Rechen-Institut, Zentrum f\"ur Astronomie der
Universit\"at Heidelberg, Germany; cassan@ari.uni-heidelberg.de}

\altaffiltext{5}{Laboratoire d'Astrophysique de Bordeaux (CNRS; Universit\'e Bordeaux I), BP 89, F-33270 Floirac, France; wakelam@obs.u-bordeaux1.fr}

\altaffiltext{6}{Institut d'astrophysique de Paris, CNRS (UMR 7095),
Universit\'e Pierre \& Marie Curie, France; beaulieu@iap.fr}

\altaffiltext{7}{The HOLMES collaboration}

\begin{abstract}
{We study the formation conditions of icy planetesimals in protoplanetary disks in order to determine the composition of ices in small and cold extrasolar planets. Assuming that ices are formed from hydrates, clathrates, and pure condensates, we calculate their mass fractions with respect to the total quantity of ices included in planetesimals, for a grid of disk models. We find that the composition of ices weakly depends on the adopted disk thermodynamic conditions, and is rather influenced by the initial composition of the gas phase. {The use of a plausible range of molecular abundance ratios and the variation of the relative elemental carbon over oxygen ratio in the gas phase of protoplanetary disks, allow us to apply our model to a wide range of planetary systems. Our results can thus be used to constrain the icy/volatile phase composition of cold planets evidenced by microlensing surveys, hypothetical ocean-planets and carbon planets, which could be detected by \Corot\ or \Kepler.}}
\end{abstract}


\section{Introduction}
\label{sec:intro} 

{Recent discoveries of low-mass extrasolar planets demonstrate the current capability to detect very low-mass planets (see e.g. Rivera et al. 2005; Beaulieu et al. 2006; Udry et al. 2007). In particular, the microlensing technique has an unequaled potential to reveal planetary companions of a few \Mearth\ with ground-based telescopes, down to a fraction of \Mearth\ when operating from space (Bennett \& Rhie 2002). Using the PLANET network of ground-based  telescopes, Beaulieu et al.\ (2006) reported the detection of OGLE~2005-BLG-390Lb (hereafter, OGLE~390Lb), a 5.5-\Mearth\ planet at 2.6~\AU\ from a faint M star. Although the microlensing planet detection efficiency is $\sim$50 times higher for Jovian-mass than for sub-Uranus-mass planets (Cassan \& Kubas 2007; Kubas et al. 2007), detections are equally distributed, which supports the core-accretion theory prediction that low-mass planets should be more common than gas giants. Models of OGLE~390Lb can thus apply to a progressively unveiled, large population of planets (Ehrenreich \& Cassan 2007). Ehrenreich et al.\ (2006a) argued that such cold objects ($T_{{\rm surf}}\sim$40~K) could host a subsurface ocean under an ice shell for several billion years. Besides, there is a growing interest in modeling low-mass exoplanets, as illustrated by the descriptions of the possible internal structures and atmospheres of `super-Earths' (Valencia et al.\ 2007; Sotin et al.\ 2007; Ehrenreich et al.\ 2006b) or `ocean-planets' (Kuchner 2003; L\'eger et al.\ 2004; Sotin et al.\ 2007; Selsis et al.\ 2007).}

These low-mass planets could originate from beyond the snow line of their
protoplanetary disks and later migrate closer to the star. Kuchner (2003) and
L\'eger et al.\ (2004) suggested they should consequently be composed by a
large fraction of volatile species, typically $\sim$50\% in mass, including
not only H$_2$O, but also `minor' compounds, brought in by icy planetesimals
during the formation of the planet. Such compounds are carbon monoxide (CO),
carbon dioxide (CO$_2$),  {methanol (CH$_3$OH)}, methane (CH$_4$), ammonia
(NH$_3$), molecular nitrogen (N$_2$), or hydrogen sulfide (H$_2$S).

This work aims at constraining the composition of the condensed volatile-phase, or cryo\-sphere\footnote{The cryosphere includes all volatile `layers' condensed in the planet: the ice shell, the potential subsurface ocean, and the icy mantle; cryo ($\kappa\rho\acute{\upsilon}o$) simply means cold.} of small and cold planets. We examine the primordial conditions leading to the formation of icy planetesimals eventually building those planets. The composition of ices is a key parameter in planetary internal structure and thermal evolution models. `Minor' species determine the structure of cold planet icy mantles, the presence of liquid layers (oceans), and the atmospheric composition. For instance, the existence of subsurface oceans on Europa, Ganymede, or Titan, depends on the composition of the volatiles incorporated in these moons{\footnote{{The presence of ammonia in the interiors of major icy satellites lowers the melting temperature of water and leads to the preservation of deep liquid layers during their thermal history (see e.g. Spohn \& Schubert 2003; Mousis \& Gautier 2004).}}. More generally, the composition of ices in protoplanetary bodies is a function of the intricated properties of the protoplanetary disk: the formation distance of planetesimals to the star, its initial gas phase composition, its thermodynamic evolution and the carbon-to-oxygen relative abundances (C:O).

In \S~\ref{sec:disks}, we use a generic accretion disk model (Papaloizou \& Terquem 1999; Alibert et al.\ 2005) to represent the plausible thermodynamic conditions of the nebula from which the planetesimals were produced. We then examine the gas phase chemistry that may occur in such a disk and derive the resulting abundances of the main volatile species in  \S~\ref{sec:evolution}. In \S~\ref{sec:comp}, we determine the composition of ices in the planetesimals produced in the cooling protoplanetary disk. Finally, \S~\ref{sec:implications} is devoted to the discussion of the implications for exoplanets detected or to be detected with the microlensing and transit techniques. {We finally discuss in  \S~\ref{sec:implications} the broader implications of our results for the composition of planets detected and to be detected by the microlensing and transit techniques.}

\section{Models of protoplanetary disks}
\label{sec:disks}

The composition of ices incorporated in planetesimals is determined by their condensation sequence in the cooling protoplanetary disk. The thermodynamic conditions of the formation of a condensate at a given distance to the star are determined by the intersection of its stability curve (as shown in Fig.~\ref{cooling}; see \S~\ref{sec:comp} for details) with the evolutionary track of the disk (hereafter cooling curve) calculated at the appropriate location. {The cooling curve for a given radial distance results from the determination of the disk vertical structure which, in turn, derives from the turbulent model used in this work. While a full description of our turbulent model of accretion disk can be found in Papaloizou \& Terquem (1999) and Alibert et al.\ (2005), we simply provide in this section a concise outline of the underlying assumptions.}\\

Assuming cylindrical symmetry in the protoplanetary disk, the vertical structure along the $z$-axis at a given radial distance $r$ is calculated by solving a system of three equations, namely, the equation for hydrostatic equilibrium, the energy conservation, and the diffusion equation for the radiative flux (eqs.~(1), (2), and (3) in Alibert et al.\ 2005, respectively). The system variables are the local pressure $P(r,z)$, temperature $T(r,z)$, density $\rho(r,z)$, and viscosity $\nu(r,z)$. {Following Shakura \& Sunyaev (1973), the turbulent viscosity $\nu$ is expressed in terms of Keplerian rotation frequency $\Omega$ and sound velocity $C_s$, as $\nu = \alpha C_s^2 / \Omega$, where $\alpha$ is a parameter characterizing the turbulence in the disk.} If the disk evolution is mainly governed by viscosity and not by other processes (e.g., photoevaporation by the central star or nearby stars), the lower $\alpha$ is, the slower the disk evolves, and inversely. In this approximation, the disk temperature, pressure and surface
density are determined once the stellar mass $M_\star$, the radial distance $r$, and the viscosity parameter $\alpha$ have been given. At a given $r$, the temporal evolution of $P$ and $T$ draws the cooling curves, shown in Fig.~\ref{cooling} for $\alpha = 2 \times 10^{-3}$, $M_\star = 0.22$~\Msun,
and $r = 2.6$ and $10$~\AU, values representative of the OGLE~390Lb system.\\

{In fact, OGLE~390Lb (Beaulieu et al. 2006) is an example of such a frozen planetary system, whose physical parameters ($M_\star$ = 0.22$+2.1 \atop -1.1$ \Msun~and semi-major axis $a$ = 2.6$+1.5 \atop -0.6$ \AU) makes it relevant for our study. Taking into account the uncertainties on the OGLE~390Lb system parameters, we consider protoplanetary disks around stars of masses 0.11, 0.22, and 0.43 \Msun. Furthermore, we consider three regimes of disk turbulence, with values of $\alpha$ equal to  $1 \times 10^{-4}$, $2 \times 10^{-3}$, and $1 \times 10^{-2}$, thus yielding nine disk models.} For each model, two extreme planetary formation scenarios are considered: \emph{in situ} formation at the present-day location of the planet (2, 2.6, and 4.1~\AU, depending on the choice of $M_\star$), and formation with large scale migration, where we consider the composition of planetesimals at 10~\AU\ from the star. These two extreme cases should bracket the actual formation processes.\\

It is important to mention that, after having performed calculations using the afore-mentioned range of values for the parameters $\alpha$ and $M_\star$ of our disk model, we found that the volatile trapping conditions (temperature, pressure and density of gas) remain almost constant at any distance to the star, whatever the adopted value of these parameters. These statements imply that, whatever the input parameters adopted when modelling the disk, and regardless of the formation location considered for icy planetesimals beyond the present-day position of the planet, their composition (in wt\%) remains almost constant (consistent within $\sim$0.5\%), provided that the gas-phase abundances are homogeneous in the disk. In the following, we choose to fix $M_\star = 0.22$~\Msun, $\alpha = 2 \times 10^{-3}$, and $r = 2.6$~\AU~in all our calculations.

\section{Evolution of the gas phase in protoplanetary disks}
\label{sec:evolution}

\subsection{Distribution of oxygen, carbon and nitrogen between refractory and volatile components}
\label{sec:gas}

{In order to define the gas phase composition in protoplanetary disks, we consider both refractory and volatile components. Refractory components include rocks and organics. According to Lodders (2003), rocks contain $\sim$23\% of the total oxygen in the nebula. The fractional abundance of organic carbon is assumed to be 55\% of total carbon (Pollack et al.\ 1994; Sekine et al.\ 2005), and the ratio of C:O:N included in organics is supposed to be 1:0.5:0.12 (Jessberger et al.\ 1988; Sekine et al.\ 2005). We then assume that the remaining O, C, and N exist only under the form of H$_2$O, CO$_2$, CO, CH$_3$OH, CH$_4$, N$_2$, and NH$_3$.  Hence, once the gas phase abundances of elements are defined, the abundances of CO, CO$_2$,  CH$_3$OH, CH$_4$, N$_2$ and NH$_3$ are determined from the adopted CO:CO$_2$:CH$_3$OH:CH$_4$ and N$_2$:NH$_3$ gas phase molecular ratios given in \S~\ref{sec:chemistry}, and from the C:O:N relative abundances set in organics. Once the abundances of these molecules are fixed, the remaining O gives the abundance of H$_2$O. 

We follow this strategy for the determination of the composition of ices produced from a disk owning a solar elemental composition (see \S~\ref{sec:solarcomp}). In the case of a disk where gas phase abundances of all elements but carbon remain solar (see \S~\ref{sec:notsolarcomp}), we consider the variation of the C:O ratio in the gas phase. Indeed, as the C:O ratio increases, all the available O goes into organics, CH$_3$OH, CO, and CO$_2$, so that the gas phase becomes H$_2$O free and the remaining C is in the form of CH$_4$. This approach is consistent with C:O-dependent chemical models of dense interstellar clouds from which ISM ices are formed (Watt 1985). Similarly, at higher values of the C:O ratio, the abundances of NH$_3$, N$_2$, CO$_2$, CO, and CH$_3$OH progressively decrease as the amounts of oxygen and nitrogen increase in organics, since their abundances relative to carbon remain fixed with C:O:N = 1:0.5:0.12 in these components.}

\subsection{Gas phase chemistry}
\label{sec:chemistry}

{Here, we investigate the plausible range of abundance ratios that may exist between main volatile species in the gas phase of protoplanetary disks.} The temperature in the disk models is initially high enough to sublimate all ISM ices entering the disk and, in particular, in the formation zone of planetesimals that will eventually take part to the planet accretion. Thus, similarly to the case of the Solar nebula (Mousis et al. 2002), the gas phase molecular ratios in the protoplanetary disks are presumed to derive directly from those in ISM where the gas and solid phases coexist. However, in some cases, gas phase chemistry may affect these molecular ratios in the disks. In particular, taking example of the equilibrium calculations of the distribution of sulfur in the Solar nebula (Pasek et al. 2005), we assume that this element is present in protoplanetary disks in the form of H$_2$S and other refractory sulfurated components. We then set H$_2$S:H$_2$ = 0.7 $\times$ (S:H$_2$)$_\sun$ (Mousis et al.\ 2006 and references therein). 

In addition, the value of the N$_2$:NH$_3$ ratio is quite uncertain in the gas phase of our disk models, although current chemical models of the ISM predict that molecular nitrogen should be much more abundant than ammonia (Irvine \& Knacke 1989). Indeed, the N$_2$:NH$_3$ ratio may have been much lower in the disks since it has been shown that the conversion of N$_2$ into NH$_3$ can be accelerated by the catalytic effect of local Fe grains in the Solar nebula (Lewis \& Prinn 1980; Fegley 2000). We assume that disks considered here behave like the Solar nebula. In all our following calculations, we consider three cases, namely N$_2$:NH$_3$ = 0.1:1, 1:1 and 10:1 in the gas phase.

On the other hand, following Prinn \& Fegley (1981,1989), the net reactions relating CO, CH$_4$ and CO$_2$ in a gas dominated by H$_2$ are

\begin{equation}
\mathrm{CO + H_2O \leftrightharpoons CO_2 +H_2}
\label{eq_chim1}
\end{equation}

\begin{equation}
\mathrm{CO + 3H_2 \leftrightharpoons CH_4 +H_2O}
\label{eq_chim2}
\end{equation}

\noindent which all proceed to the right with decreasing temperature at constant pressure. Reaction (\ref{eq_chim1}) has been studied by Talbi \& Herbst (2002) who demonstrated that its rate coefficient is negligible, even at temperatures as high as 2000 K (of the order of $\sim$4.2~$\times~10^{-22}$~cm$^3$~s$^{-1}$). Such a high temperature range is only reached at quite close distances to the star and at early epochs only in our protoplanetary disks models. As a result, the amount of carbon species produced through this reaction is insignificant during the whole lifetime of the protoplanetary disks considered in this work. Reaction (\ref{eq_chim2}) has been studied by Lewis \& Prinn (1980) and Mousis et al. (2002) in the case of the Solar nebula. These authors found that the initial CO:CH$_4$ ratio was not significantly modified throughout the Solar nebula, except quite close to the Sun. Since the protoplanetary disks considered here are less massive and colder than the Solar nebula model used by Mousis et al. (2002), we estimate that the CO:CH$_4$ ratio is also poorly affected in the gas phase of our models. Finally, we assume that the CH$_3$OH:CH$_4$ gas phase ratio present in the protoplanetary disks is similar to that in ISM. Note also that ISM ices contain a small fraction of H$_2$CO. Unfortunately, since there exist no data on the stability curves of H$_2$CO either in the form of clathrate or as a pure condensate, this molecule cannot be considered in the gas phase composition of our protoplanetary disks. Hence, we assume that H$_2$CO has been largely hydrogenated into CH$_3$OH on interstellar grains.

From the afore-mentioned considerations, we set CO:CH$_3$OH:CH$_4$ = 70:2:1 in the gas phase of the disks, values that are consistent with the ISM measurements that consider the contributions of both gas and solid phases in the lines of sight (Frerking et al. 1982; Ohishi et al. 1992; Ehrenfreund  \& Schutte 2000; Gibb et al. 2000). In addition, we opted to test the influence of several CO$_2$:CO gas phase ratios, namely the two most abundant carbon volatile compounds, in the 0.1--1 range, whose extreme values correspond to the ones measured or expected in the gas and solid phases of ISM, respectively (Ehrenfreund  \& Schutte 2000; Gibb et al. 2004). Table \ref{Nominal} summarizes the nominal gas phase molecular ratios adopted in our model of protoplanetary disk. 

Note that the surface densities of our disk models still remain higher than 100 g cm$^{-2}$ in the formation zone of planetesimals as long as the carbon and nitrogen species considered here are not trapped in clathrates or condensed as pure ices. Such a high density implies that cosmic rays do not produce enough active ions such as H$_3^+$ or He$^+$ in the gas phase in order to convert significant amounts of CO and N$_2$ into CO$_2$, CH$_4$, NH$_3$ and HCN (see e.g. Aikawa et al. 1999).

\section{Formation of icy planetesimals}
\label{sec:comp}

The process by which volatiles are trapped in icy planetesimals, illustrated in Fig.~\ref{cooling}, is calculated using the stability curves of hydrates, clathrates and pure condensates, and the evolutionary tracks detailing the evolution of temperature and pressure at 2.6 and 10~\AU\ (see \S~\ref{sec:disks}) for CO:CO$_2$:CH$_3$OH:CH$_4$ = 70:10:2:1 and N$_2$:NH$_3$ = 1:1 in the gas phase. The corresponding gas phase abundances of the volatile compounds are given in Table \ref{lodders}. The stability curves of hydrates and clathrates derive from Lunine \& Stevenson (1985)'s compilation of laboratory data, from which data is available at relatively low temperature and pressure. On the other hand, the stability curves of pure condensates used in our calculations derive from the compilation of laboratory data given in the CRC Handbook of chemistry and physics (Lide 2002). The cooling curve intercepts the stability curves of the different ices at some given temperature, pressure and surface density conditions. For each ice considered, the domain of stability is the region located below its corresponding stability curve. The clathration process stops when no more crystalline water ice is available to trap the volatile species.

\subsection{Ices produced in a gas phase of solar composition}
\label{sec:solarcomp}

We calculate here the composition of ices produced from a gas phase of solar composition, with CO:CO$_2$:CH$_3$OH:CH$_4$ = 70:10:2:1 and N$_2$:NH$_3$ = 1:1 (our nominal gas phase molecular ratios). We obtain similar results in all our disk models: NH$_3$, H$_2$S, xenon (Xe) and CH$_4$  are entirely trapped by H$_2$O, as NH$_3$-H$_2$O hydrate, H$_2$S-5.75H$_2$O, Xe-5.75H$_2$O and CH$_4$-5.75H$_2$O  clathrates, respectively. About 60\% of CO is trapped as CO-5.75H$_2$O clathrate. The remaining CO, as well as N$_2$, krypton (Kr) and argon (Ar), whose clathration normally occurs at lower temperatures, remain in the gas phase until the disk cools enough to allow the formation of pure condensates (between 20 and 30~K). Note that during the cooling of disks, CO$_2$ is the only species crystallizing as a pure condensate before being trapped by H$_2$O to form a clathrate. Hence, pure ice of CO$_2$ is the only existing solid form containing this species. In addition, we have considered only the formation of pure ice of CH$_3$OH in our model since, to our best knowledge, no experimental data concerning the stability curve of its associated clathrate have been reported in the literature.

Using the trapping/formation conditions of the different ices calculated at a given heliocentric distance in the protoplanetary disks, and knowing their gas phase abundances, one can estimate their mass ratios with respect to H$_2$O in the accreting planetesimals. Indeed, the volatile, $i$, to water mass ratio in these planetesimals is determined by the relation given by Mousis \& Gautier (2004):         

\begin{equation}
{m_i = \frac{X_i}{X_{H_2O}} \frac{\Sigma(r; T_i, P_i)}{\Sigma(r; T_{H_2O}, P_{H_2O})}},
\end{equation}

\noindent where $X_i$ and $X_{H_2O}$ are the mass mixing ratios of the volatile $i$ and H$_2$O with respect to H$_2$ in the protoplanetary disk, respectively. $\Sigma(R; T_i, P_i)$ and $\Sigma(R; T_{H_2O}, P_{H_2O})$ are the surface density of the disk at a distance $r$ from the star at the epoch of hydratation or clathration of the species $i$, and at the epoch of condensation of water, respectively. From ${\it m_i}$, it is possible to determine the mass fraction $M_i$ of species $i$ with respect to all the other volatile species taking part to the formation of an icy solid:

\begin{equation}
{M_i = \frac{m_i}{\displaystyle \sum_{j=1,n} m_j}},
\end{equation}

\noindent with $\displaystyle \sum_{i=1,n} M_i = 1$.

\noindent The resulting composition of ices in the forming planetesimals does not depend on the chosen disk model (see \S~\ref{sec:disks}), but rather on the postulated composition of the gas phase. 

Figure \ref{comp1} shows the variation of the composition of icy planetesimals formed in the disks as a function of the CO$_2$:CO ratio (shown for values between 0.1:1 and 1:1), with CO:CH$_3$OH:CH$_4$ = 70:2:1 and N$_2$:NH$_3$ = 1:1 in the gas phase and Table \ref{fractions} gives the mass ratios of the ices for the two extreme cases CO$_2$:CO = 0.1:1 and 1:1. The composition of ices given in Fig. \ref{comp1} is then valid for solids formed at any distance within the protoplanetary disks calculated in \S~\ref{sec:disks}, provided that the gas phase is homogeneous and the disks are initially warm enough at that location to vaporize the ices falling in from the ISM. It can be seen that water remains the most abundant ice in mass, whatever the value of the CO$_2$:CO gas phase ratio adopted in the disks. Interestingly enough, CO is the main carbon species trapped within planetesimals for CO$_2$:CO $\leq$ 0.4 in the gas phase. At higher CO$_2$:CO ratios, CO$_2$ becomes the main carbon species incorporated in solids. We also note that the relative amounts of H$_2$O and of other carbon species decrease as the CO$_2$:CO ratio increases. In contrast, the mass fractions of NH$_3$, N$_2$, H$_2$S, and of the noble gases are weakly influenced by the variation of the CO$_2$:CO gas phase ratio.\\ 

Figures \ref{comp2}--\ref{comp3} show the variation of the composition of icy planetesimals formed in the protoplanetary disks as a function of the CO$_2$:CO ratio in the same gas phase conditions as in Fig. \ref{comp1}, but with N$_2$:NH$_3$ = 0.1:1 and 10:1, respectively. It can be seen that, whatever the initial N$_2$:NH$_3$ ratio adopted in the gas phase, the mass fractions of other volatile compounds is weakly influenced by this parameter. Hence, an extension of the range of variation of this ratio is not warranted.

\subsection{Ices produced in a carbon- or oxygen-rich gas phase}

\label{sec:notsolarcomp} 

The composition of ices forming in a disk with a given C:O ratio is obtained assuming that gas phase abundances of all elements but carbon remain solar. {In order to test different values of the C:O ratio, which may differ from the `protosolar' value ((C:O)$_\odot$ $\sim$0.5; Lodders 2003), we define a carbon-to-oxygen enrichment factor, $f \equiv$ (C:O)$_\mathrm{disk}$ / (C:O)$_\odot$. Indeed,} Kuchner \& Seager (2005) suggested that extrasolar systems can form from carbon-rich clouds where $f > 2$, while observations of the debris disk around $\beta$~Pic hint toward an extremely carbon-rich post-protoplanetary environment ($f$ $\sim$18; Roberge et al.\ 2006). On the other hand, ices may have crystallized in protoplanetary disks formed from oxygen-rich clouds. For instance, observations of the Orion bar yielded $f$ $\sim$0.5 (Walmsley et al.\ 1999). {In the present study, we cover the afore-mentioned range of observed values by varying $f$ between 0.1 and 20.}

Figure~\ref{fig:CsurO} shows the composition of ices formed in protoplanetary disks as a function of $f$. The condensation sequence of ices is calculated as in \S~\ref{sec:solarcomp} and the values shown here for $f = 1$ match those found in \S~\ref{sec:solarcomp} for  CO:CO$_2$:CH$_3$OH:CH$_4$ = 70:10:2:1 and N$_2$:NH$_3$ = 1 (our nominal gas phase composition). For low $f$ (0.1 to 1.6), H$_2$O is the major constituent of ices (86 to 34\% in mass, respectively). For $1.6 < f < 3.0$, the main ice in planetesimals is CO ($\geq 34\%$ in mass, with a peak at 53\% in mass for $f = 2$). For $f \geq 3$, CH$_4$ becomes the main icy compound ($\geq 33\%$ in mass).

\section{Summary and discussion}
\label{sec:implications}

{In this work, we have quantified the `bulk' amount of minor species included in icy planetesimals produced in protoplanetary disks. The composition of ices calculated here can be translated into the composition of planetary ices. However, it should be regarded as an \emph{initial composition}, and the individual species mass fractions as \emph{upper limit values}. 

The main assumptions that allowed us to determine the composition of icy planetesimals are recalled below:

\begin{enumerate}[-]
\item Taking into account the uncertainties on the physical parameters of the OGLE 390Lb system, and using an evolutionary turbulent model,  we have calculated a range of plausible disks thermodynamic conditions from which low-mass extrasolar planets may have formed;

\item Ices are formed from hydrates, clathrates and pure condensates produced during the cooling of the protoplanetary disks. The efficiency of clathration is ruled by the amount of available crystalline water ice in the disks;

\item A fraction of the available oxygen in disks is included in rocks. Similarly, organics include a part of the O, C and N initial budget available in the gas phase. Finally, the remaining O, C, and N are distributed under the form of H$_2$O, CO$_2$, CO, CH$_3$OH, CH$_4$, N$_2$, and NH$_3$;

\item Assuming solar elemental abundances in the initial gas phase, plausible ranges of CO:CO$_2$:CH$_3$OH:CH$_4$ and N$_2$:NH$_3$ gas phase molecular ratios have been employed in disks to investigate the variation of icy solids composition;

\item In some cases, C:O has been varied in order to constrain the composition of icy planetesimals forming from carbon- or oxygen-rich nebulae.

\end{enumerate}

\noindent From these hypotheses, the main results derived from our calculations are the following:

\begin{enumerate}[-]
\item  The composition of ices weakly depends on the adopted disk thermodynamic conditions, and is rather influenced by the initial composition of the gas phase;

\item Using solar gas phase elemental abundances, H$_2$O is dominant regardless of the adopted CO$_2$:CO molecular ratio;

\item The variation of the adopted N$_2$:NH$_3$ molecular ratio weakly influences the mass fractions of other volatiles incorporated in ices;

\item Planets formed in carbon-rich environments are depleted in water and would be essentially CO- or CH$_4$-rich.

\end{enumerate}}

So far, models of volatile-rich low-mass extrasolar planets have considered almost pure-H$_2$O cryospheres. On the other hand, L\'eger et al.\ (2004) estimated that hypothetical ocean-planets should contain some `minor' volatile species, with molecular abundances similar to that of cometary ice in the Solar system. However, we have to bear in mind that there are some uncertainties in determining the composition of comets. In particular, it is difficult to directly detect molecules in cometary nuclei. Instead, only molecules and radicals (daughter molecules), which are formed in the comae via photolysis, are observed. From these observations, the molecular composition of the parent molecules is determined using chemical reaction models. Hence, the molecular composition of the parent molecules is somewhat model dependent. Moreover, it is unclear whether the cometary nuclei are chemically homogeneous, and how much they were reprocessed in their formation stages and during their stay in the Oort cloud or in the Kuiper belt (Aikawa et al. 1999). There are indeed several evidences indicating the presence of macroscopic internal heterogeneity in cometary nuclei. For example, observations of comet Halley showed that the relative production rate of major volatiles varied more than tenfold, and outburst were observed for some given species, such as CO$_2$-rich outbursts (Mumma et al. 1993). Thus, comets cannot be considered as good indicators of planetary ices composition.

\paragraph{Comparison with Solar System.}
In the framework of giant planets formation scenarios, we point out that similar calculations reproduce the current abundances of volatile species in the atmospheres of Jupiter and Saturn (Mousis \& Marb\oe uf 2006). Moreover, they can explain the current composition of Titan's atmosphere measured by the
\emph{Huygens} probe (Alibert \& Mousis 2007). Estimating the composition of icy bodies devoid of atmosphere is less straightforward since the spectroscopy measurements are actually limited to their surface. Indeed, despite of the detections of H$_2$O, CO, CO$_2$, CH$_4$, NH$_3$, and others, at the surface of several icy satellites (Brown \& Cruikshank 1997; Hibbitts et al. 2002,2003), we do not have, to our knowledge, a real insight of their `bulk' composition. As a result, internal structure models of these bodies usually consider H$_2$O icy mantles and set the mass fractions of minor species as free parameters (Sohl et al. 2002; Spohn \& Schubert 2003).

\paragraph{Implications for Cold Low-Mass Extrasolar Planets.} 
Ehrenreich et al.\ (2006a) reported that cold low-mass extrasolar planets could be described as frozen ocean-planets, with a subsurface liquid H$_2$O layer surviving between low- and high-pressure ice shells, in a large version of a Ganymede-like internal structure model. The heat source allowing the existence of the liquid layer is radiogenic energy from within the underlaying rocky mantle. {As this heat source decreases with time, the subsurface ocean eventually freezes in a few Gyr, as predicted by Ehrenreich et al. (2006a). However, the freezing might occur on a longer timescale if the cryosphere of the planet contains small amounts of NH$_3$ since ammoniated water has a lower freezing point than pure H$_2$O. Hence, a subsurface ocean of ammoniated water can exist in the range where H$_2$O is the major component of ice for $f \leq 1.6$ solar, while the mass fraction of NH$_3$ ice remains nearly constant around $\sim$3.7--4.3\% (see Fig.~\ref{fig:CsurO}).}

\paragraph{Implications for `Hot Ocean-Planets'.} Our results can also be used to constrain the icy/volatile phase composition of warmer planets that formed beyond the snow line and migrated closer to the star. This is the case of ocean-planets described by Kuchner (2003), L\'eger et al.\ (2004), and Selsis et al.\ (2007). They noted that the presence of CO$_2$ in ocean-planets could dramatically change their cooling history and structure, because CO$_2$ would maintain the atmosphere in a hot state, hence preventing the condensation of an icy mantle and ocean. A small fraction of CO$_2$ in the ices is sufficient to trigger this effect, because given the ice-to-rock mass fraction of an ocean-planet ($\sim$1), it represents in absolute several hundred times the quantity of CO$_2$ in the atmosphere of Venus. Indeed, we predict that {$\sim$7 \%} of the icy mass is initially made of CO$_2$, assuming a solar-composition in our disk models, with a peak at 25\% for $f = 2$. This could lead to increase the predicted mean density of an ocean-planet, making it more difficult to discriminate ocean-planets from super-Earths in planetary candidates yielding from \Corot\ and \Kepler\ transit surveys (Selsis et al.\ 2007). \textit{An ocean-planet forming in a carbon-rich environment could rather be a `carbon planet' (see below)}.

\paragraph{Implications for `Carbon Planets'.}
Kuchner \& Seager (2005) conjectured up planets forming in carbon-rich environments. `Carbon planets' could be forming for $f > 1.6$ in the gas phase. Such environments could include carbon-rich stellar surrounds, like the debris disk around $\beta$~Pic (Roberge et al.\ 2006). The main refractory material composing these hypothetical planets should be silicon carbide (SiC), rather than metals and silicates usually considered for the refractory phase of planets forming at solar C:O ratio. Our present work allows to constrain the volatile phase of such planets. The major ice compound is CO, with mass fractions {between $\sim$34 to 53\% for $1.6 \leq f \leq 3.0$ in the gas phase conditions adopted in \S~3.2.} In this case and according to Fig.~\ref{fig:CsurO}, we expect CO, CO$_2$, and CH$_4$ to be the other main compounds as $f$ increases. It is interesting to note that these planets should be almost completely dry, as water amounts to a negligible value. For even larger values of $f$, CH$_4$ becomes the most abundant ice in carbon planets, by far {($\geq 80 \%$ in mass for $f \geq 5$)}.

\acknowledgements
UM, OM, AC and JPB acknowledge the financial support of the ANR HOLMES. DE acknowledges support of the ANR `Exoplanet Horizon 2009'. AC and OM acknowledge the ESO Visitor Programme for a financing one-month stay at ESO Santiago in March 2007, where part of the discussions were carried out. We thank Franck Selsis, Jean-Marc Petit and an anonymous Referee for useful remarks and suggestions.

%
\clearpage
\begin{table}
\caption[]{Nominal Molecular ratios adopted in the gas phase of protoplanetary disk.}
\begin{center}
\begin{tabular}{ll}
\hline
\hline
\noalign{\smallskip}
Molecular ratio    & value\\	
\noalign{\smallskip}
\hline
\noalign{\smallskip}
CO:CH$_4$   		   	& 70  \\
CO$_2$:CH$_4$      	& 10  \\
CH$_3$OH:CH$_4$    	& 2  \\
N$_2$:NH$_3$       		& 1  \\
\hline
\end{tabular}
\begin{flushleft}
\end{flushleft} 
\end{center}
\label{Nominal}
\end{table}


\clearpage
\begin{table}
\caption[]{Gas phase abundances (molar mixing ratio with respect to H$_2$) of major species in the Solar nebula from Lodders (2003) with CO:CO$_2$:CH$_3$OH:CH$_4$ = 70:10:2:1 and N$_2$:NH$_3$ = 1:1 (our nominal gas phase composition).}
\begin{center}
\begin{tabular}{lclc}
\hline
\hline
\noalign{\smallskip}
Species X &  X/H$_2$  & species X  & X/H$_2$ \\	
\noalign{\smallskip}
\hline
\hline
\noalign{\smallskip}
O		& $7.33 \times 10^{-4}$		& N$_2$		& $4.05 \times 10^{-5}$ \\
C		& $2.62 \times 10^{-4}$ 		& NH$_3$   	& $4.05 \times 10^{-5}$ \\
N   		& $1.22 \times 10^{-4}$    		& S        		& $3.66 \times 10^{-5}$ \\
H$_2$O  	& $4.43 \times 10^{-4}$   		& Ar       		& $8.43 \times 10^{-6}$ \\
CO      	& $2.21 \times 10^{-4}$    		& Kr       		& $4.54 \times 10^{-9}$ \\	
CO$_2$  	& $3.16 \times 10^{-5}$   		& Xe       		& $4.44 \times 10^{-10}$ \\
CH$_4$  	& $3.16 \times 10^{-6}$   		& CH$_3$OH  	& $6.31 \times 10^{-6}$  \\ 	  
H$_2$S    &	$2.56 \times 10^{-5}$       &	                                     \\
\hline
\end{tabular}
\end{center}
\label{lodders}
\end{table}

\clearpage
\begin{table}
\caption[]{Ratio of the mass of ice ${\it i}$ to the global mass of ices (wt\%) in planetesimals formed in protoplanetary disks, calculated for CO$_2$:CO = 0.1:1 and 1:1 in the vapor phase. Both ratios are calculated with CO:CH$_3$OH:CH$_4$~=~70:2:1 and N$_2$:NH$_3$~=~1:1 in the gas phase.}
\begin{center}
\begin{tabular}{lcc}
\hline
\hline
\noalign{\smallskip}
Species &  CO$_2$:CO = 0.1:1 & CO$_2$:CO = 1:1  \\
\noalign{\smallskip}
\hline
\noalign{\smallskip}
H$_2$O		& $6.19 \times 10^{-1}$ 		& $4.70 \times 10^{-1}$	\\
CO$_2$   		& $5.16 \times 10^{-2}$ 		& $2.93 \times 10^{-1}$	\\
CO   	   		& $1.98 \times 10^{-1}$   		& $1.14 \times 10^{-1}$	\\
CH$_4$   		& $2.25 \times 10^{-3}$	  	& $1.24 \times 10^{-3}$ 	\\
CH$_3$OH 	& $1.28 \times 10^{-2}$		& $7.06 \times 10^{-3}$	\\
NH$_3$   		& \multicolumn{2}{c}{$3.88 \times 10^{-2}$} \\
H$_2$S   		& \multicolumn{2}{c}{$4.74 \times 10^{-2}$} \\
N$_2$    		& \multicolumn{2}{c}{$2.27 \times 10^{-2}$} \\
Ar 			& \multicolumn{2}{c}{$6.26 \times 10^{-3}$}  \\
Kr 			& \multicolumn{2}{c}{$1.04 \times 10^{-5}$}  \\
Xe  			& \multicolumn{2}{c}{$2.65 \times 10^{-6}$}  \\	
\hline
\end{tabular}
\end{center}
\label{fractions}
\end{table}

\clearpage
\begin{figure}
\resizebox{\hsize}{!}{\includegraphics[angle=-90]{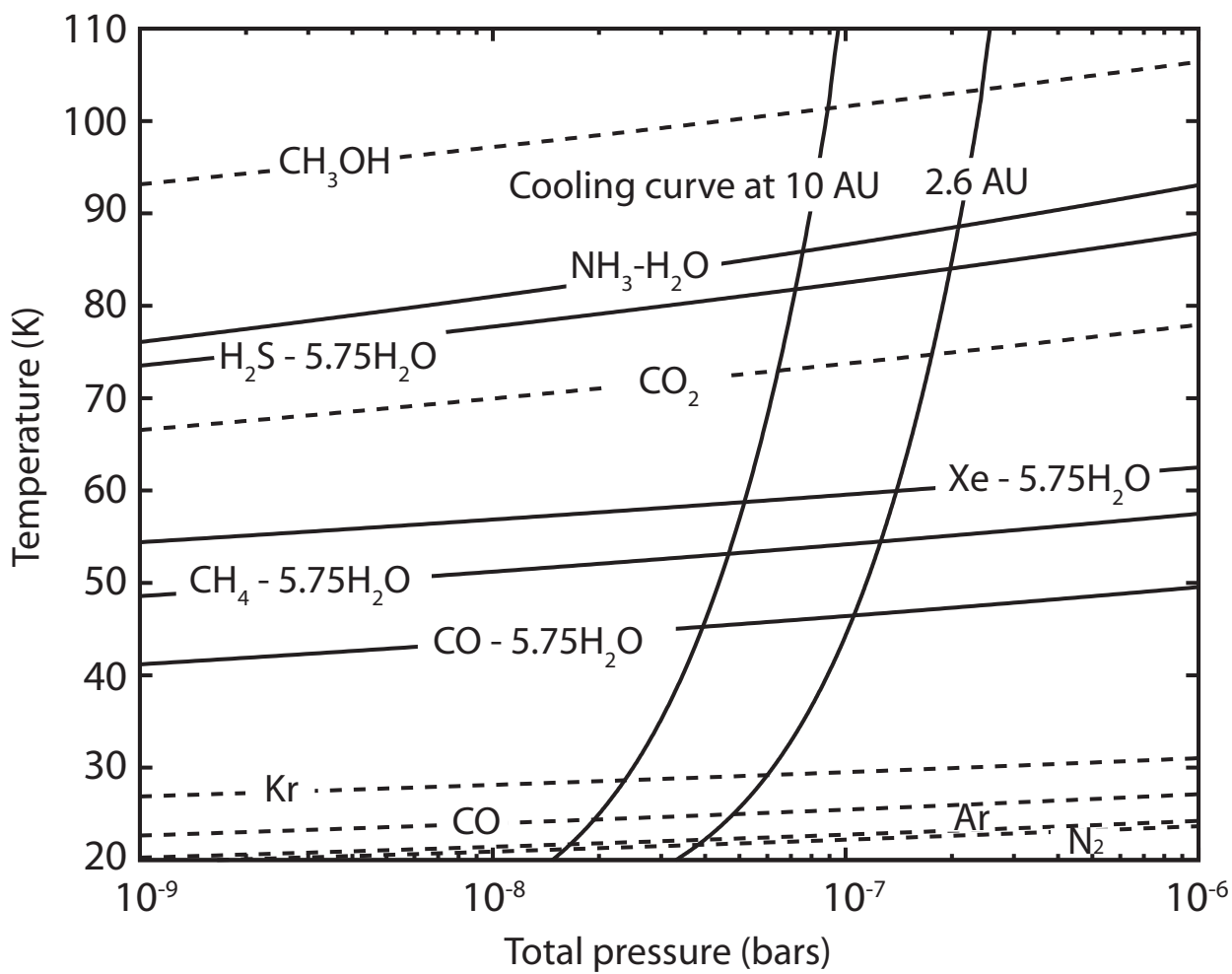}} \caption{Stability
curves of hydrate (NH$_3$-H$_2$O), clathrates ($X$-5.75H$_2$O)
[\emph{solid lines}], and pure condensates [\emph{dotted lines}], and cooling
curves of the protoplanetary disk model at two given distances to the star, 2.6
and 10~\AU\ (the evolution of the disk proceeds from high to low temperatures).
Abundances of elements are solar, with CO:CO$_2$:CH$_3$OH:CH$_4$ = 70:10:2:1 and N$_2$:NH$_3$ = 1:1 in the gas phase (our nominal gas phase composition). Species remain in the gas phase above the stability curves. Below, they are trapped as clathrates or simply condense.} 
\label{cooling}
\end{figure}

\clearpage
\begin{figure}
\resizebox{\hsize}{!}{\includegraphics[angle=-90]{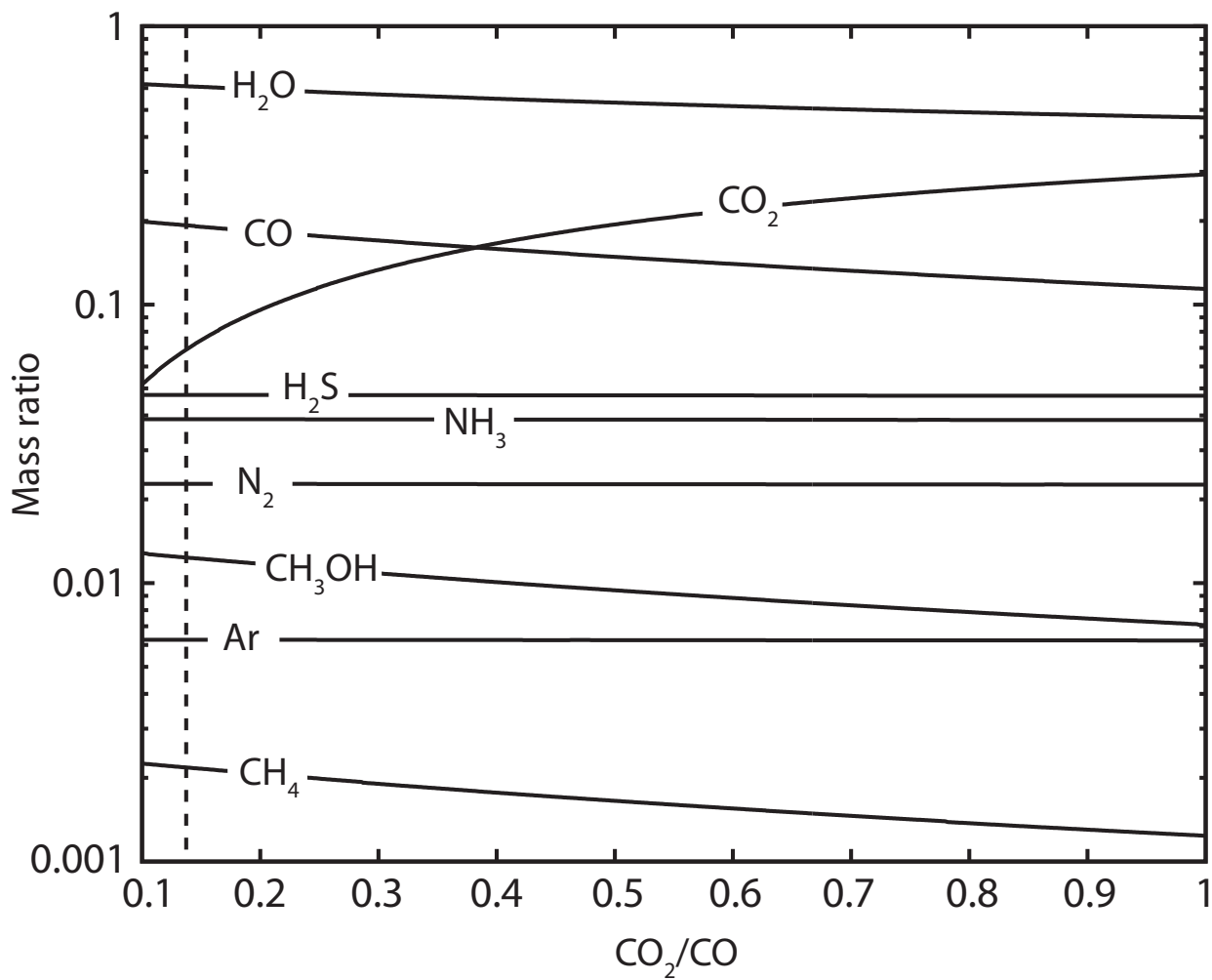}} \caption{Composition of ices
(wt\%) incorporated in planetesimals produced in a disk where the abundances of
all elements are solar, as a function of the adopted CO$_2$:CO ratio in the
initial gas phase. The composition is expressed in mass ratio (mass of ice $i$
to the total mass of ices). Abundances of considered elements are solar with CO:CH$_3$OH:CH$_4$
 = 70:2:1 and N$_2$:NH$_3$ = 1:1 in the gas phase. The intersection of the vertical dashed line with the other curves gives the mass fraction ratios of ices formed from CO:CO$_2$:CH$_3$OH:CH$_4$ = 70:10:2:1 and N$_2$:NH$_3$ = 1:1, namely our nominal gas phase composition.} 
\label{comp1}
\end{figure}

\clearpage
\begin{figure}
\resizebox{\hsize}{!}{\includegraphics[angle=-90]{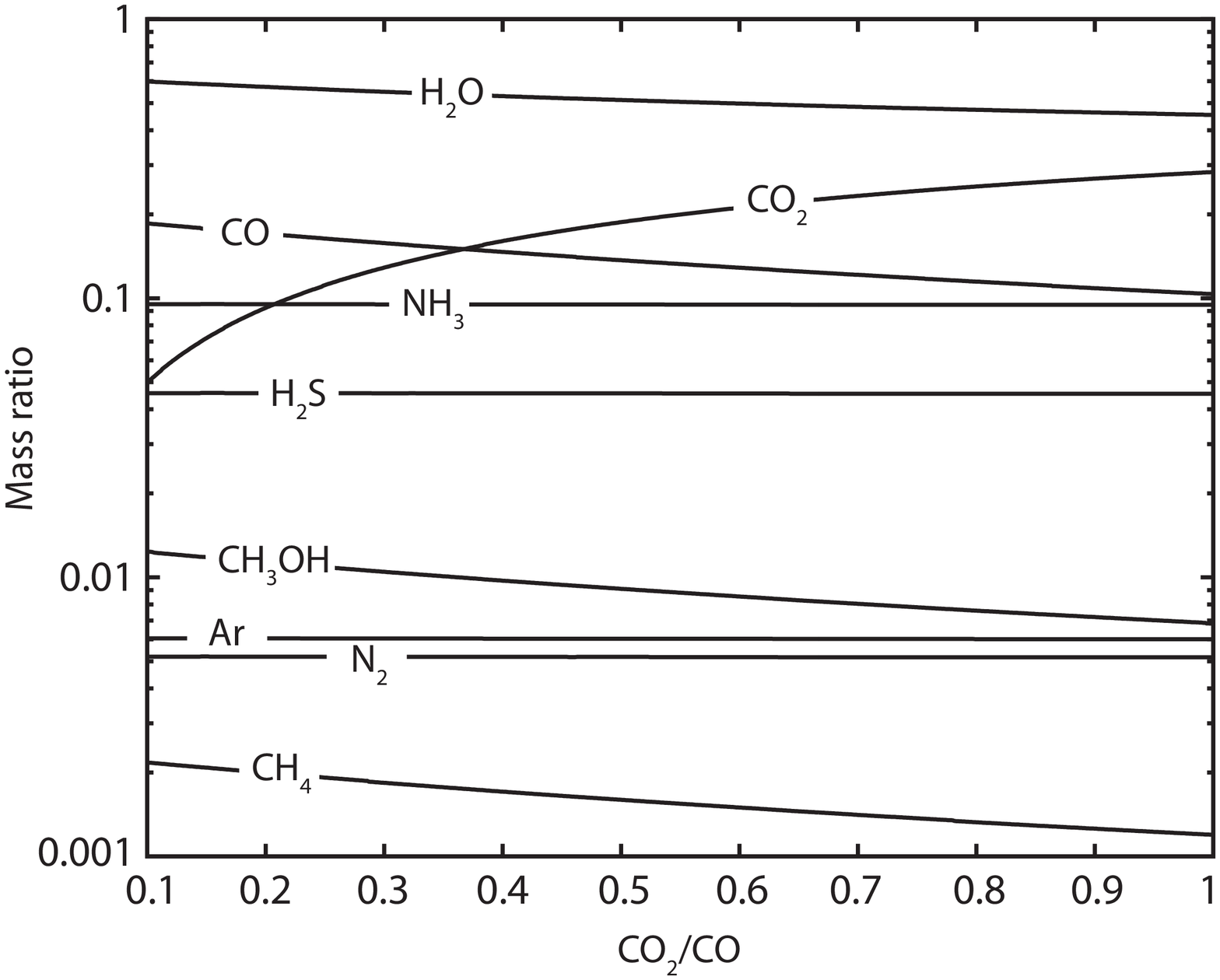}} \caption{Same as Fig. \ref{comp1} but for N$_2$:NH$_3$ = 0.1:1 in the gas phase.} 
\label{comp2}
\end{figure}

\clearpage
\begin{figure}
\resizebox{\hsize}{!}{\includegraphics[angle=-90]{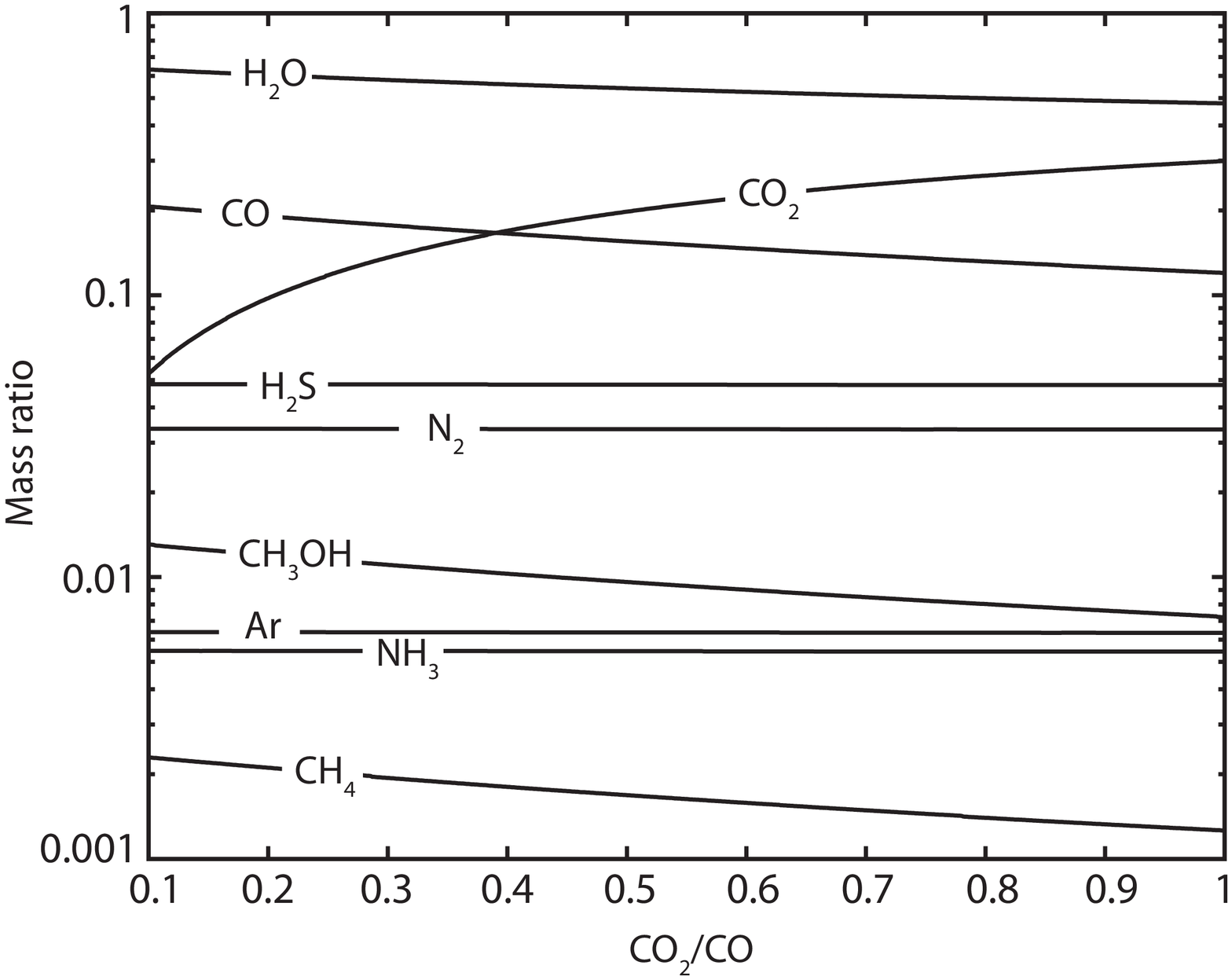}} \caption{Same as Fig. \ref{comp1} but for N$_2$:NH$_3$ = 10:1 in the gas phase.} 
\label{comp3}
\end{figure}

\clearpage
\begin{figure}
\resizebox{\hsize}{!}{\includegraphics[angle=-90]{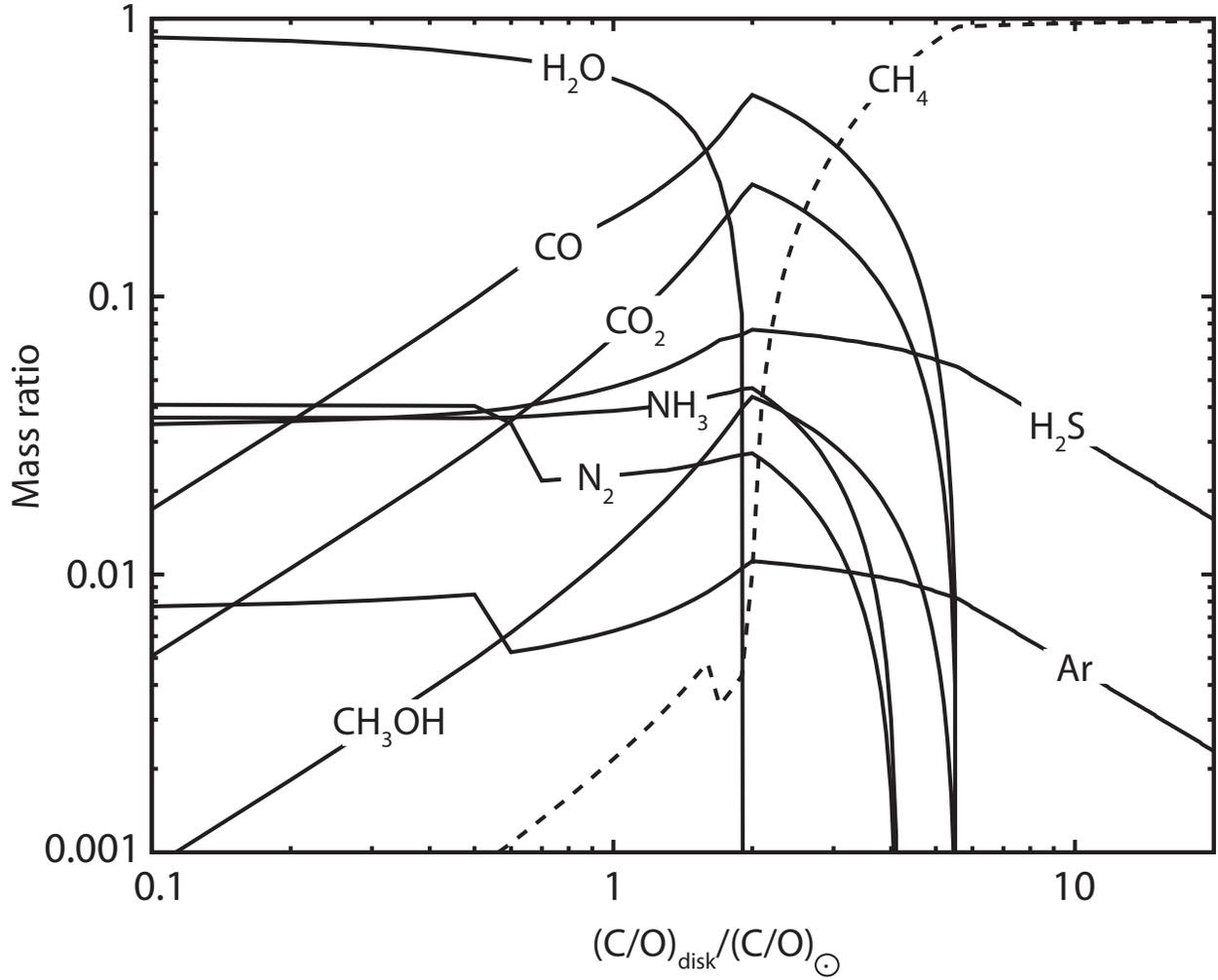}} \caption{Composition of ices (wt\%) in planetesimals produced in a disk where all elements but carbon are solar, as a function of the C/O enrichment factor $f \equiv \mathrm{(C/O)_\mathrm{disk}/\mathrm{(C/O)}_\odot}$ in the initial gas phase.}
\label{fig:CsurO}
\end{figure}

\end{document}